\documentclass[showpacs,10pt,twocolumn,prb]{revtex4}
\usepackage{amsmath}
\usepackage{amssymb}
\usepackage{graphics}
\usepackage{epsfig}
\usepackage{color}

\begin{document}

\title{Anisotropic magnetism, resistivity, London penetration depth and
magneto-optical imaging of superconducting K$_{0.80}$Fe$_{1.76}$Se$_{2}$
single crystals}
\author{R. Hu, K. Cho, H. Kim, H. Hodovanets, W. E. Straszheim, M. A.
Tanatar, R. Prozorov, S. L. Bud'ko, P. C. Canfield}
\affiliation{Ames Laboratory, U.S. DOE and Department of Physics and Astronomy, Iowa
State University, Ames, IA 50011, USA}
\date{\today }

\begin{abstract}
Single crystals of K$_{0.80}$Fe$_{1.76}$Se$_{2}$ were suscessfully grown
from a ternary solution. We show that although crystals form when cooling a
near stoichiometric melt, crystals are actually growing out of a ternary
solution that remains liquid to at least 850 $^{o}C$. We investigated their
chemical composition, anisotropic magnetic susceptibility and resistivity,
specific heat, thermoelectric power, London penetration depth and flux
penetration via magneto-optical imaging. Whereas the samples appear to be
homogeneously superconducting at low temperatures, there appears to be a
broadened transtion range close to $T_{c}$ $\sim $ 30 K that may be
associated with small variations in stociometry.
\end{abstract}

\maketitle

The iron-based superconductors have attracted intense research attention
because of their high transition temperature and their possibly
unconventional pairing mechanism, correlated to magnetism.\cite{Kenji}$^{-}$%
\cite{Johnpierre} Similar to cuprate superconductors, iron-based
superconductors have layered structures; the planar Fe layers tetrahedrally
coordinated by As or chalcogen anions (Se or Te) are believed to be
responsible for superconductivity. Stacking of the FeAs building blocks with
alkali, alkaline earth or rare earth oxygen spacer layers forms the basic
classes of iron arsenic superconductors in these compounds: 111-type AFeAs%
\cite{Wang}, 122-type AFe$_{2}$As$_{2}$\cite{Rotter}$^{-}$\cite{Jasper},
1111-type ROFeAs\cite{Kamihara}$^{,}$\cite{Chen} and more complex block
containing phases, e.g. Sr$_{2}$VO$_{3}$FeAs\cite{Zhu}, Sr$_{3}$Sc$_{2}$Fe$%
_{2}$As$_{2}$O$_{5}$\cite{Zhu2}, Sr$_{4}$Sc$_{2}$Fe$_{2}$As$_{2}$O$_{6}$.%
\cite{Chen2} The simple binary 11-type iron chalcogenide has no spacer
layers and superconductivity can be induced by doping FeTe with S\cite%
{Rongwei} or Se.\cite{Mizu} Different from the other iron-based
superconductors, FeSe is a superconductor\cite{Hsu}, T$_{c}\sim 8$ K,\ with
no static magnetic order and its transition temperature can be increased up
to 37 K by applying pressure\cite{Med} or 15 K in FeSe$_{0.5}$Te$_{0.5}$.%
\cite{Mizu} More recently, superconductivity above 30 K has been reported in
A$_{x}$Fe$_{2-y}$Se$_{2}$ (A = K, Cs, Rb or Tl)\cite{Guo}$^{-}$\cite{Fang},
a compound with the same unit cell structure as the AFe$_{2}$As$_{2}$
compounds. These new compounds generally have a width of formation, show
strong dependence of electrical transport properties on its stoichiometry/Fe
vacancy and are in very close proximity to an insulating state.\cite{Fang}$%
^{,}$\cite{Wangdm} The growth of single crystals of K$_{x}$Fe$_{2-y}$Se$_{2}$
has been reported in a number of publications using various claimed growth
methods: self-flux growth\cite{Ying}, Bridgeman method.\cite{Wangdm} Due to
the off-stoichiometric nature of the K$_{x}$Fe$_{2-y}$Se$_{2}$, wide ranges
of the values of $x,y$ ($0.6\leq x<1$ and $0\leq y\leq 0.59$)\cite{Ying}$%
^{-} $\cite{Zava} have been reported for the superconducting crystals with
similar $T_{c}$ values ($\sim 31-33$ $K$) from several groups. This raises
the question what the correlation between superconductivity and
stoichiometry is, if there is any, and whether there is a uniformity problem
with the single crystal samples. Thus well controlled samples are needed and
it is desirable to check the homogeneity of the superconducting crystals and
understand their growth.

In this work, we will try to clarify the growth details and present
elemental analysis, anisotropic magnetization and resistivity data, as well
as measurements of heat capacity, thermoelectrical power, London penetration
depth and flux penetration on K$_{0.80}$Fe$_{1.76}$Se$_{2}$ single crystals.

Single crystals of K$_{x}$Fe$_{2-y}$Se$_{2}$ were first grown from K$_{0.8}$%
Fe$_{2}$Se$_{2}$ melt, as described in Ref. 20. First the FeSe precursor was
prepared by reacting stoichiometric Fe and Se at 1050 $^{o}C$. Then K and
FeSe with a nominal composition of K$_{0.8}$Fe$_{2}$Se$_{2}$ were placed in
an alumina crucible that was sealed in an amorphous silica tube. Due to
potassium attack on the silica tube, this primary ampoule was sealed into a
secondary, larger silica tube to prevent exposure to air if the first
ampoule degraded enough to crack. The growth was placed in a furnace in a
vented enclosure and heated to 1050 $^{o}C$, where it was held for a 2 hours
soak. The furnace temperature was then and slowly lowered to 750 $^{o}C$
over 50 hours; the furnace was then turned off and the sample "furnace
cooled" over an additional 10 hours. Once the ampoules were opened, large ( $%
\sim 1\times 1\times 0.02$ $cm^{3}$) dark shiny crystals could be
mechanically separated from the solidified melt. The crystals are moderately
air-sensitive and should be handled under an Ar atmosphere.

The above growth procedure clearly is not simply the cooling of a
stoichiometric melt to form a congruently melting, line compound. There is
clear loss of K from the melt (as seen by the attack of the inner ampoule)
and there is a clear mixed phase resultant sample, consisting of the desired
single crystalline phase separated by fine polycrystalline material. In
order to better establish the nature of the growth of K$_{x}$Fe$_{2-y}$Se$%
_{2}$ the above procedure was repeated for a starting composition of KFe$%
_{3} $Se$_{3}$. The sample was heated to 1050 $^{o}C$, held for 2 hours and
slowly cooled to 850 $^{o}C$ at which point the remaining solution was
decanted. The resulting crystals were about 1/2 the area and thickness of
the the samples cooled to 750 $^{o}C$, but they were well formed and no
longer embedded in solidified flux. This result clearly shows that K$_{x}$Fe$%
_{2-y}$Se$_{2}$ crystals are grown out of a ternary melt.

\begin{figure}[tbp]
\centerline{\includegraphics[scale=0.35]{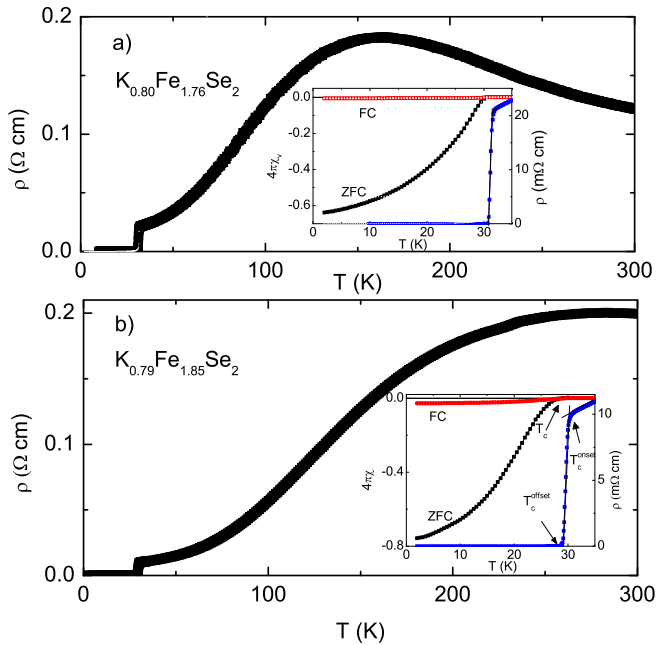}} \vspace*{-0.3cm}
\caption{Comparison of the in-plane resistivity, and low temperature
magnetic susceptibility of two types of K$_{x}$Fe$_{2-y}$Se$_{2}$ single
crystals, a) furnace cooled; b) decanted sample. Inset shows the low
temperature region of the resistivity (to the right axis) together with
zero-field-cooled and field-cooled magnetic susceptibility in a field of 50
Oe.}
\end{figure}

Crystals were characterized by powder x-ray diffraction using a Rigaku
Miniflex X-ray diffractometer. The actual chemical composition was
determined by wavelength dispersive x-ray spectroscopy (WDS) in a JEOL
JXA-8200 electron microscope. Magnetic susceptibility was measured in a
Quantum Design MPMS, SQUID magnetometer. In plane AC resistivity $\rho _{ab}$
was measured by a standard four-probe configuration. Measurement of $\rho
_{c}$ was made in the two-probe configuration. Contacts were made by using a
silver alloy. For $\rho _{c}$, contacts were covering the whole \textit{ab }%
plane area.\cite{Jeffrey} Thermoelectrical power\ measurements were carried
out by a dc, alternating temperature gradient (two heaters and two
thermometers) technique.\cite{Mun} Heat capacity data were collected using a
Quantum Design PPMS. The in-plane London penetration depth was measured by
using a tunnel-diode resonator(TDR) oscillating at 14 MHz and at temperature
down to 0.5 K.\cite{RuslanTDR} Magneto-optical imaging was conducted by
utilizing the Faraday effect in bismuth-doped iron garnet indicators with
in-plane magnetization.\cite{Doro} A flow-type liquid $^{4}$He cryostat with
sample in vacuum was used. The sample was positioned on top of a copper cold
finger and an indicator was placed on top of the sample. The cryostat was
positioned under polarized-light reflection microscope and the color images
could be recorded on video and high-resolution CCD cameras. When linearly
polarized light passes through the indicator and reflects off the mirror
sputtered on its bottom, it picks up a double Faraday rotation proportional
to the magnetic field intensity at a given location on the sample surface.
Observed through the (almost) crossed analyzer, we recover a 2D image.\cite%
{Joos}

The x-ray diffraction pattern can be indexed using space group I4/mmm. The
lattice parameters refined by Rietica were $a=3.8897(8)\mathring{A}$ and $%
c=14.141(3)\mathring{A}$. They are in good agreement with the previous
reported values in Ref. 20 ($a=3.8912\mathring{A}$, $c=14.139\mathring{A}$),
but disagree with Ref. 19 ($a=3.9136(1)\mathring{A}$, $c=14.0367(7)\mathring{%
A}$) and Ref. 27 ($a=3.9034\mathring{A}$, $c=14.165\mathring{A}$), in
lattice constant \textit{c. }It is probably due to the different
stoichiometry of the crystals.

Previous reported stoichiometries of K$_{x}$Fe$_{2-y}$Se$_{2}$ crystals were
determined by the semi-quantitative Energy Dispersive X-ray (EDX)
spectroscopy.\cite{Ying}$^{-}$\cite{Zava} Here we performed precise
measurement of the stoichiometry using WDS. Twelve measurement spots were
spread uniformly across the crystal surfaces of dimension approximately $%
3\times 3$ $mm^{2}$. All of the spots showed consistent results. By
averaging 12 spots, the stoichiometry was determined to be $%
K:Fe:Se=0.80(2):1.76(2):2.00(3)$ for the crystal grown from solidified melt
and $K:Fe:Se=0.79(2):1.85(4):2.00(4)$ for the crystal grown from solution,
where the atomic numbers of K and Fe are normalized to two Se per formula
unit and the standard deviation $\sigma $ is taken as the compositional
error and shown in parentheses after value. The spread of composition, the
difference between the maximum and minimum values of the measurements, is
0.07, 0.06 and 0.10 for K, Fe and Se respectively for crystal grown from
solidified melt and 0.04, 0.12 and 0.09 for crystal grown from solution,
roughly within $3\sigma $ of a normal distribution of random variable. The
crystals grown from solution have very similar composition to the furnace
cooled samples, with only a little higher concentration of Fe.

Basic, temperature dependent electrical resistivity and magnetization
measurements were performed on crystals grown by both the furnace cooled and
decanted methods. The in-plane resistivity of the furnace cooled sample is
very similar to that of earlier reports.\cite{Ying}$^{,}$\cite{Wangdm} There
is a broad resistive maxima centered near 160 K followed by a lower
temperature drop by nearly a factor of 6 ($\rho (300K)/\rho (35K)$). There
is a sharp transition to a zero resistance state. The inset to Fig. 1a shows
the low temperature resistivity as well as the in-plane, magnetic
susceptibility (H=50 Oe). The superconducting transition temperature, $%
T_{c}=30.1$ $K$, can be inferred by the first deviation of the
zero-field-cooled curve from normal magnetic susceptibility. It is
consistent with the $T_{c}^{offset}=30.9$ $K$, inferred from resistivity.
The transition is sharp with a width of 0.7 K and $T_{c}^{onset}=31.6$ K.

The in-plane resistivity of single crystals grown out of solution (the
decanted samples) is shown in Fig. 1b. It exhibits a broad maximum around
280 K and becomes superconducting below 30 K. The inset shows the in-plane
magnetic susceptibility ($H=50$ $Oe$) and resistivity at low temperature.
Superconducting transition temperature, $T_{c}=29.0$ $K$, can be inferred by
the first deviation of the zero-field-cooled curve from normal magnetic
susceptibility. It is consistent with the $T_{c}^{offset}=29$ $K$, inferred
from resistivity. The transition is sharp with a width of 1 K and $%
T_{c}^{onset}=30.0$ K. The temperature of the broad resistive peak in Fig.
1b is higher than the one shown in Fig. 1a and the Tc value is slightly
lower. Wang \textit{et al.} showed that the position of the hump is
sensitive to Fe deficiency.\cite{Wangdm} With decreasing Fe deficiency, the
hump shifts to higher temperature. This may imply that our decanted crystal
has a slightly higher Fe concentration, reasonable for a crystal grown out
of solution with a greater excess of Fe-Se. Even with these slight
differences, the WDS analysis and the data shown in figure 1 demonstrate
that these are closely related compositions with very similar properties.
Given the somewhat larger size of the furnace cooled samples, as well as
their similarity to samples from earlier reports, for the rest of this paper
we will focus their fuller characterization. 
\begin{figure}[tbp]
\centerline{\includegraphics[scale=0.6]{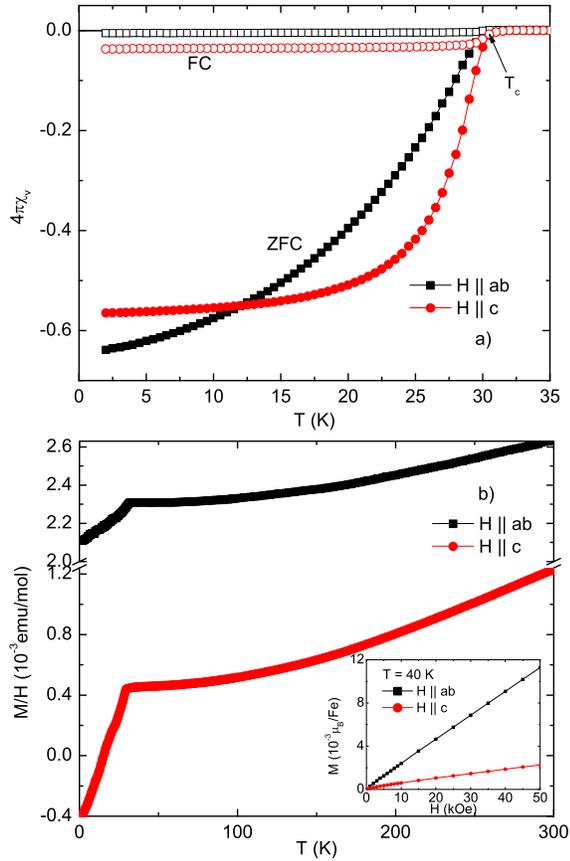}} \vspace*{-0.3cm}
\caption{a) Temperature dependence of low field (H = 50 Oe) magnetic
susceptibility for H$\Vert $\textit{ab} and H$\Vert $\textit{c}; b) Magnetic
susceptibility M/H, measured in 50 kOe for two field directions. Inset shows
field dependence of magnetization at 40 K for both field directions.}
\end{figure}

\begin{figure}[tbp]
\centerline{\includegraphics[scale=0.4]{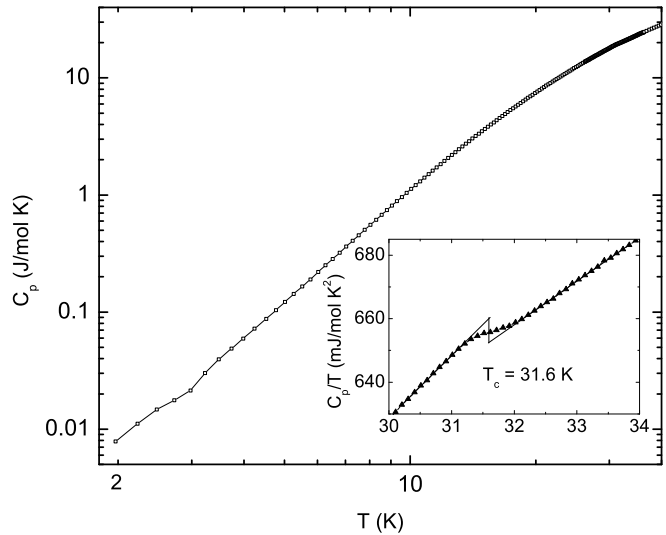}} \vspace*{-0.3cm}
\caption{Heat capacity as a function of temperature on a log-log plot. Inset
shows the heat capacity jump at the superconducting transition. The solid
line is an isoentropic estimate of $T_{c}$ and $\Delta C_{p}$.}
\end{figure}

Figure 2a shows the magnetic susceptibility of K$_{0.80}$Fe$_{1.76}$Se$_{2}$
for two directions of an applied field of 50 Oe. For magnetic field along 
\textit{c} axis, a correction of demagnetization for a thin rectangular
sample has been made. For H $\Vert $ \textit{ab,} the zero-field-cooled
(ZFC) curve decreases slowly with temperature and for H $\Vert $ \textit{c}
the transition becomes sharper. Similar behavior can be seen in Tl$_{0.58}$Rb%
$_{0.42}$Fe$_{1.72}$Se$_{2}$.\cite{Wanghd} This temperature dependence of
the ZFC curve is similar to an inhomogeneous superconductor with a range of
transition temperatures and may be related to the small spread of
stoichiometry found in WDS data. Both of the zero-field-cooled (ZFC) curves
in Fig. 2a approach -0.6 consistent with bulk superconductivity and $T_{c}$
inferred from both curves is the same, $T_{c}=30.1\pm 0.1$ K, within
experimental error.

The magnetic susceptibility M/H (H = 50 kOe) as a function of temperature
for both field directions is shown in Fig. 2b. Similar temperature
dependence is observed for both field directions, i.e. M/H decreases almost
linearly with decreasing temperature above 150 K and shows a sudden drop
below 30 K associated with superconductivity. $\chi _{ab}$ is clearly larger
than $\chi _{c}$ over the whole temperature range. The inset to Fig. 2b
shows the magnetization as a function of magnetic field at $T=40$ $K$.
Magnetization is linear with magnetic field for both directions. It also
should be noted that the magnitude of magnetization, even at highest field,
is very small, $10^{-3}\mu _{B}/Fe$. This indicates that there are no
ferromagnetic impurities or Curie-Weiss like, local Fe moments and the
system might be deep in an antiferromagnetic state, similar to what was
suggested for Cs$_{0.8}$Fe$_{2}$Se$_{1.96}$\cite{Sher} and K$_{0.8}$Fe$_{1.6}
$Se$_{2}$\cite{Weibao}, or in a non-magnetic state.

\begin{figure}[tbp]
\centerline{\includegraphics[scale=0.6]{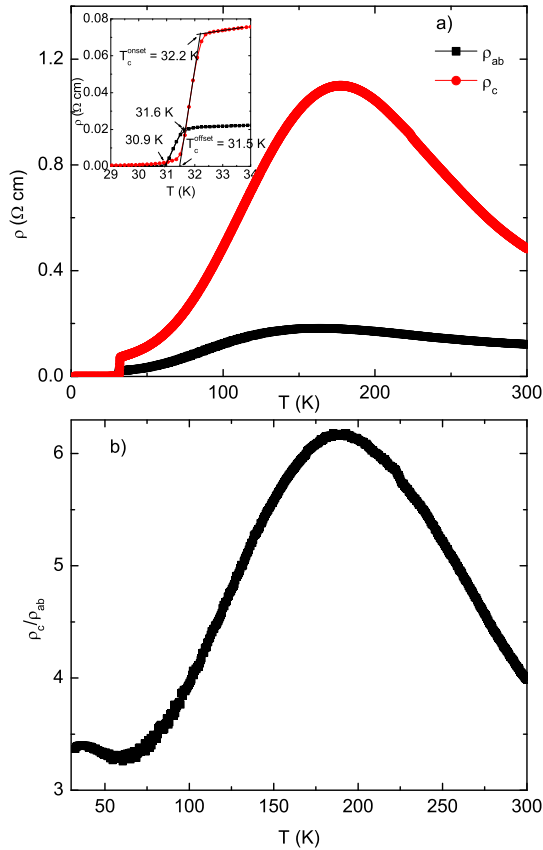}} \vspace*{-0.3cm}
\caption{a) Anisotropic resistivity as a function of temperature. Inset is
an expanded view around the transition. b) Anisotropy of resistivity vs
temperature.}
\end{figure}
\begin{figure}[tbp]
\centerline{\includegraphics[scale=0.35]{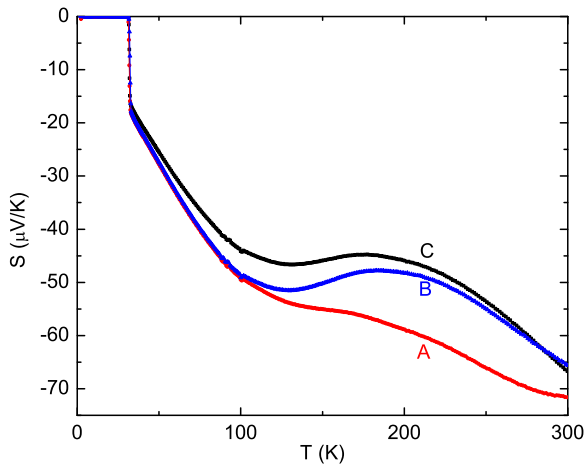}} \vspace*{-0.3cm}
\caption{Thermoelectric power as a function of temperature. Samples A and B
use silver paste as contact (contact resistance $\sim 1-3$ $k\Omega $).
Sample C uses silver wires attached by In-Sn solder as contact (contact
resistance $\sim 200$ $\Omega $).}
\end{figure}

Heat capacity data was collected to verify the bulk thermodynamic nature of
the superconducting transition. $C_{p}$ vs $T$ at low temperature is shown
in Fig. 3 on a log-log plot. In the superconducting state, below 15 K, $C_{p}
$ roughly follows a $T^{3}$ power law. This implies a dominant phonon
contribution and a very small electronic term. $C_{p}/T$ vs T is plotted in
the inset for $T\sim T_{c}$ and a clear jump of heat capacity associated
with the superconducting transition at 31.6 K is seen and $\Delta C_{p}/T=7.7
$ $mJ/mol$ $K^{2}$, can be identified. If we assume the value of normal
state electronic heat capacity coefficient $\gamma _{n}=5.8$ $mJ/mol$ $K^{2}$
(as in Ref. 38), $\Delta C_{p}/\gamma _{n}T=1.33$ of K$_{0.80}$Fe$_{1.76}$Se$%
_{2}$ is close to the weak coupling BCS value and is in variance with the
strong coupling conclusion in Ref. 38. On the other hand, this more likely
implies that a reliable conclusion about the coupling strength can not be
made due to the difficulty of estimating the normal state electronic
contribution $\gamma _{n}$.

Anisotropic resistivity as a function of temperature is shown in Fig. 4a. It
is clear that there is a broad maximum peak around 160 K for $\rho _{ab}$
and 180 K for $\rho _{c}$. The difference of maximum positions suggest that
they result from a crossover rather than transition. The temperature range
of this broad maxima does not correlate with any anomalies in magnetic
susceptibility. The anisotropy is probably due to the layered structure of K$%
_{0.80}$Fe$_{1.76}$Se$_{2}$. Figure 4b shows the anisotropy $\rho _{c}/\rho
_{ab}$, reaches the maximum of 6 around 180 K and decreases to 4 around 300
K. It is comparable to the anisotropy of AFe$_{2}$As$_{2}$.\cite{Tanatar}
But a much larger resistivity anisotropy of 30-45 was reported in (Tl,K)Fe$%
_{x}$Se$_{2}$\cite{Wanghd}, this implies that the specific composition
influences carrier tunneling significantly. An expanded view around the
superconducting transition is shown in the inset to Fig. 4a. For both of the
current directions, the transition width is about 0.7 K, but $T_{c}$ for $%
\rho _{c}$ is slightly higher than that of $\rho _{ab}$. The transition
temperatures for the two current directions are very close to the one
inferred from the heat capacity measurement using an isoentropic
construction, as well as resistivity and susceptibility measurements.

\begin{figure}[tbp]
\centerline{\includegraphics[scale=0.9]{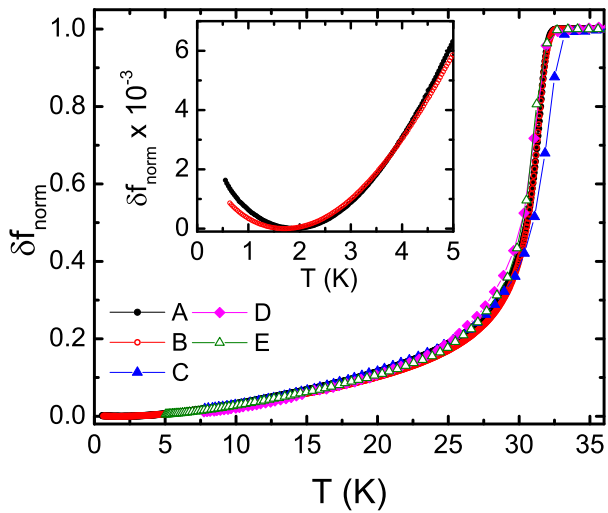}} \vspace*{-0.3cm}
\caption{Normalized London penetration depth expressed via resonant
frequency shift, $\Delta f_{norm}=(f(T)-f(T_{c}))/(f(T_{c})-f(T_{min}))$
proportional to magnetic susceptibility. $f(T_{min})$ is the resonant
frequency at the lowest temperature $\simeq $ 0.5 K. $f(T_{c})$ is the
frequency in the normal state right above $T_{c}$. Inset shows an upturn,
presumably due to paramagnetic ions and/or impurities below 2 K from two
samples A and B. }
\end{figure}

The thermoelectrical power (TEP) as a function of temperature is shown in
Fig. 5. We present results for three different samples: for samples A and B
silver paste was used for electrical and thermal contact, for sample C
silver wires were soldered to the sample by In-Sn solder and then electrical
/ thermal contact was established between the wires and the contact pads by
silver paste . For all three samples $T_{c}$ inferred from $S(T)=0$ is 31.6
K, consistent with all of our previous measurements. The data for three
samples are similar in the whole temperature range. The origin of local
minimum and maximum in $100-200$ K is not clear, but is very likely to be
associated with the multiband structure of K$_{0.80}$Fe$_{1.76}$Se$_{2}$ and
the crossover (metal-like at low temperature observed in resistivity.
Negative sign of thermopower indicates that electron like carriers are
dominant, thus in agreement with the observation of electron only pockets at
the Fermi surface by ARPES.\cite{Qian} The large absolute value of $S$ above
50 K is noteworthy and is consistent with high normal state resistivity.

London penetration depth measurements with good reproducibility were
performed on several single crystal samples. In order to compare between the
samples, we plot in Fig. 6 normalized frequency shift, proportional to
differential magnetic susceptibility, $\delta
f_{norm}=(f(T)-f(T_{c}))/(f(T_{c})-f(T_{min}))$, where $f(T_{min})$ is the
resonant frequency at the lowest temperature $\simeq $ 0.5 K and $f(T_{c})$
is the frequency in the normal state right above $T_{c}$. The samples show
consistent behavior indicating little or no variation within the batch. The
transition itself is quite unusual - it shows quite a sharp onset, but then
is smeared almost over the entire temperature interval. In this work, it
might be due to off-stoichiometry of iron and/or impurities. In principle,
it is also possible that the observed behavior is indicative of strongly
anisotropic gap function or even nodes, but we cannot rule out simple
variation of the superconducting properties as a cause of such unusual
superconducting transition in K$_{0.80}$Fe$_{1.76}$Se$_{2}$. In addition,
there is a clear upturn at low temperatures. It has been shown on both,
high-Tc cuprates\cite{Ruslan0} and 1111 pnictides\cite{Martin} that this
upturn is caused by the paramagnetic ions.

\begin{figure}[tbp]
\centerline{\includegraphics[scale=0.8]{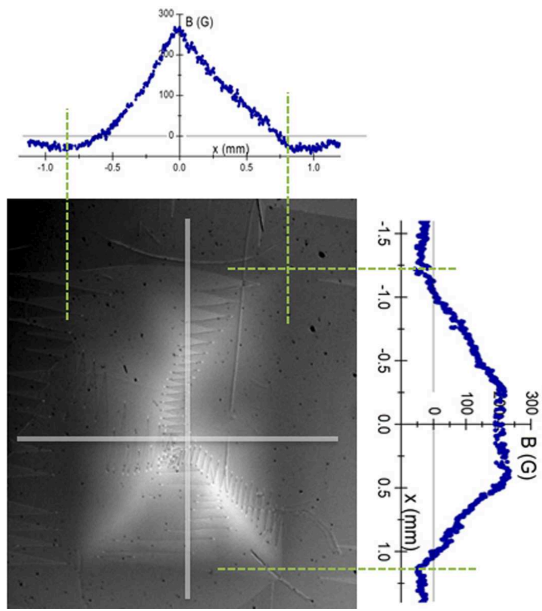}} \vspace*{-0.3cm}
\caption{Magneto-optical image of single crystal K$_{0.80}$Fe$_{1.76}$Se$%
_{2} $}
\end{figure}

To gain further insight into the homogeneity of the superconducting state
and, roughly, estimate critical current density, we performed
magneto-optical imaging. A magneto-optical image of a trapped flux is shown
in Fig. 7. In the experiment the sample was cooled in a 2 kOe magnetic field
from 40 K to 5 K. We did not observe any noticeable Meissner expulsion,
similar to other 122 pnictides.\cite{Ruslan1} When magnetic field was turned
off, it revealed a typical \textquotedblleft Bean\textquotedblright\ roof,
again similar to other pnictide superconductors.\cite{Ruslan2}$^{,}$\cite%
{Ruslan3} As can be seen in Fig. 7, magnetic flux distribution is quite
uniform and is definitely consistent with the bulk superconducting nature of
the material. However, some macroscopic variations (upper left corner) might
indicate some smooth variation of stoichiometry across the sample and may
help to explain the broadened transition curves. In order to quantify the
critical state, Fig. 7 also shows profiles of the magnetic induction taken
along two lines (shown in the figure). The remanence reaches about 250 Oe. A
simple one-dimensional estimate, using%
\begin{equation*}
\frac{4\pi }{c}j_{c}=\frac{dB}{dx}
\end{equation*}

gives:

\begin{equation*}
j_{c}=\frac{250}{0.77}\frac{10}{4\pi }\approx 2.6\times 10^{3}\text{ }%
A/cm^{2}
\end{equation*}

Of course, this estimate is very crude, but shows that the current samples
cannot support large critical current density even at low temperatures.
Similar numbers are estimated from the magnetization measurements.
Nevertheless, magneto-optical imaging is consistent with bulk
superconducting nature of K$_{0.80}$Fe$_{1.76}$Se$_{2}$ and shows that it is
not filamentary or phase separated, but rather shows smooth variation of the
stoichiometry.

In summary, single crystals of K$_{x}$Fe$_{2-y}$Se$_{2}$ have been grown via
two related methods. In both cases $T_{c}$ $\sim $ 30 K, with the furnace
cooled crystals having K$_{0.80}$Fe$_{1.76}$Se$_{2}$ composition and $%
T_{c}=30$ K (from magnetization) and decanted crystals having composition K$%
_{0.83}$Fe$_{1.86}$Se$_{2.09}$ and $T_{c}=29$ K (from magnetization). We
found moderate anisotropy in both magnetic susceptibility and electrical
resistivity with $\chi _{ab}/\chi _{c}\sim 2$ and $\rho _{c}/\rho _{ab}\sim 4
$ at 300 K. Broadened transitions seen in several measurements imply a small
variation of stoichiometry of the crystal, consistent with what was shown by
WDS analysis. It has also been shown that the critical current density of
the K$_{0.80}$Fe$_{1.76}$Se$_{2}$ is only on the order of $10^{3}$ $A/cm^{2}$%
, much smaller than those of FeAs superconductors.\cite{Tanatar2}

\textit{Note added. }During the preparation of this paper, a preprint was
posted on arxiv.org showing similar studies of anisotropy in electrical
transport and magnetization of \.{K}$_{x}$Fe$_{2-y}$Se$_{2}$.\cite{Cedomir}
The results of are consistent with ours.

\section{Acknowledgements}

This work was carried out at the Iowa State University and supported by the
AFOSR-MURI grant \#FA9550-09-1-0603 (R. Hu and P. C. Canfield). Part of this
work was performed at Ames Laboratory, US DOE, under contract \#
DE-AC02-07CH 11358 (K. Cho, H. Kim, H. Hodovanets, W. E. Straszheim, M. A.
Tanatar, R. Prozorov, S. L. Bud'ko and P. C. Canfield). S. L. Bud'ko also
acknowledges partial support from the State of Iowa through Iowa State
University. R. Prozorov acknowledges support from the Alfred P. Sloan
Foundation.


\bigskip

\begin{thebibliography}{99}
\bibitem{Kenji} Kenji Ishida, Yusuke Nakai, and Hideo Hosono, J. Phys. Soc.
Jpn., 78 062001 (2009).

\bibitem{Lumsden} M. D. Lumsden and A. D. Christianson, J. Phys.: Condens.
Matter, 22 203203 (2010).

\bibitem{Canfield} Paul C. Canfield and Sergey L. Bud'ko, Annual Review of
Condensed Matter Physics, 1, 27 (2010).

\bibitem{Johnpierre} Johnpierre Paglione and Richard L. Greene, Nature
Physics 6, 645 (2010).

\bibitem{Wang} X. C. Wang, Q. Q. Liu, Y. X. Lv, W. B. Gao, L. X. Yang, R. C.
Yu, F. Y. Li, C. Q. Jin, Solid State Commun. 148, 538 (2008).

\bibitem{Rotter} M. Rotter, M. Tegel, and D. Johrendt, Phys. Rev. Lett. 
\textbf{101}, 107006 (2008).

\bibitem{Sefat} Athena S. Sefat, Rongying Jin, Michael A. McGuire, Brian C.
Sales, David J. Singh, and David Mandrus, Phys. Rev. Lett. 101, 117004
(2008).

\bibitem{Ni} N. Ni, M. E. Tillman, J.-Q. Yan, A. Kracher, S. T. Hannahs, S.
L. Bud'{}ko, and P. C. Canfield, Phys. Rev. B \textbf{78,} 214515 (2008).

\bibitem{Jasper} A. Leithe-Jasper, W. Schnelle, C. Geibel, and H. Rosner,
Phys. Rev. Lett. \textbf{101}, 207004 (2008).

\bibitem{Kamihara} Y. Kamihara, T. Watanabe, M. Hirano, and H. Hosono, J.
Am. Chem. Soc. 130, 3296 (2008).

\bibitem{Chen} X. H. Chen, T. Wu, G. Wu, R. H. Liu, H. Chen, and D. F. Fang,
Nature 453, 761 (2008).

\bibitem{Zhu} X. Zhu, F. Han, G. Mu, P. Cheng, B. Shen, B. Zeng, and H. H.
Wen, Phys. Rev. B 79, 220512 (2009).

\bibitem{Zhu2} X. Zhu, F. Han, G. Mu, B. Zeng, P. Cheng, B. Shen, H. H. Wen,
Phys. Rev. B 79 024516 (2009).

\bibitem{Chen2} G. F. Chen, T. L. Xia, H. X. Yang, J. Q. Li, P. Zheng, J. L.
Luo, N. L. Wang, Supercond. Sci. Tech. 22, 072001 (2009).

\bibitem{Rongwei} Rongwei Hu, Emil S. Bozin, J. B. Warren, and C. Petrovic,
Phys. Rev. B 80, 214514 (2009).

\bibitem{Mizu} Y. Mizuguchi, Y. Hara, K. Deguchi, S. Tsuda, T. Yamaguchi, K.
Takeda, H. Kotegawa, H. Touand, Y. Takano, Supercond. Sci. Tech., 23 054013
(2010).

\bibitem{Hsu} F. C. Hsu, J. Y. Luo, K. W. The, T. K. Chen, T. W. Huang, P.
M. Wu, Y. C. Lee, Y. L. Huang, Y. Y. Chu, D. C. Yan and M. K. Wu, Proc. Nat.
Acad. Sci. 105, 14262 (2008).

\bibitem{Med} S. Medvedev, T. M. McQueen, I. Trojan, T. Palasyuk, M. I.
Eremets, R. J. Cava, S. Naghavi, F. Casper, V. Ksenofontov, G. Wortmann and
C. Felser, Nature Mater., 8 630 (2009).

\bibitem{Guo} Jiangang Guo, Shifeng Jin, Gang Wang, Shunchong Wang, Kaixing
Zhu, Tingting Zhou, Meng He and Xiaolong Chen, Phys. Rev. B 82, 180520
(2010).

\bibitem{Ying} J. J. Ying, X. F. Wang, X. G. Luo, A. F. Wang, M. Zhang, Y.
J. Yan, Z. J. Xiang, R. H. Liu, P. Cheng, G. J. Ye, X. H. Chen,
arXiv:1012.5552v1 (2010).

\bibitem{Li} Chun-Hong Li, Bing Shen, Fei Han, Xiyu Zhu, Hai-Hu Wen,
arXiv:1012.5637v2 (2010).

\bibitem{Fang} Minghu Fang, Hangdong Wang, Chiheng Dong, Zujuan Li, Chunmu
Feng, Jian Chen, H. Q. Yuan, arXiv:1012.5236 (2010).

\bibitem{Wangdm} D. M. Wang, J. B. He, T.-L. Xia, G. F. Chen,
arXiv:1101.0789v1 (2011).

\bibitem{ZhangY} Y. Zhang, L. X. Yang, M. Xu, Z. R. Ye, F. Chen, C. He, J.
Jiang, B. P. Xie, J. J. Ying, X. F. Wang, X. H. Chen, J. P. Hu, D. L. Feng,
arXiv:1012.5980v1 (2010).

\bibitem{Shein} I.R. Shein, A.L. Ivanovskii, arXiv:1012.5164 (2010).

\bibitem{Qian} T. Qian, X.-P. Wang, W.-C. Jin, P. Zhang, P. Richard, G. Xu,
X. Dai, Z. Fang, J.-G. Guo, X.-L. Chen, H. Ding, arXiv:1012.6017 (2010).

\bibitem{Mizuguchi} Yoshikazu Mizuguchi, Hiroyuki Takeya, Yasuna Kawasaki,
Toshinori Ozaki, Shunsuke Tsuda, Takahide Yamaguchi and Yoshihiko Takano,
Appl. Phys. Lett. 98, 042511 (2011).

\bibitem{Petrovic} D. A. Torchetti, M. Fu, D. C. Christensen, K. J. Nelson,
T. Imai, H. C. Lei, C. Petrovic, arXiv:1101.4967v1 (2011).

\bibitem{Zava} P. Zavalij, W. Bao, X. F. Wang, J. J. Ying, X. H. Chen, D. M.
Wang, J. B. He, X. Q. Wang, G.F Chen, P-Y Hsieh, Q. Huang, M. A. Green,
arXiv:1101.4882 (2011).

\bibitem{Jeffrey} Jeffrey E. Marchese, Matteo Cirillo, and Niels Gr$%
\varnothing $bech-Jensen, Phys. Rev. B 79, 094517 (2009).

\bibitem{Mun} Eundeok Mun, Sergey L. Bud'ko, Milton S. Torikachvili, and
Paul C. Canfield, Meas. Sci. Technol 21, 21055104(2010).

\bibitem{RuslanTDR} R. Prozorov and R.W. Giannetta, Supercond. Sci. Tech.
19, R41 (2006).

\bibitem{Doro} L. A. Dorosinskii, M. V. Indenbom, V. I. Nikitenko, Yu. A.
Ossip'yan, A. A. Polyanskii, and V. K. Vlasko-Vlasov, Physica C 203, 149
(1992).

\bibitem{Joos} Ch. Jooss, J. Albrecht, H. Kuhn, S. Leonhardt, and H.
Kronmuller, Rep. Prog. Phys. 65, 651 (2002).

\bibitem{Wanghd} Hangdong Wang, Chihen Dong, Zujuan Li, Shasha Zhu, Qianhui
Mao, Chunmu Feng, H. Q. Yuan, Minghu Fang, arXiv:1101.0462v1 (2011).

\bibitem{Sher} Z. Shermadini, A. Krzton-Maziopa, M. Bendele, R. Khasanov, H.
Luetkens, K. Conder, E. Pomjakushina, S. Weyeneth, V. Pomjakushin, O.
Bossen, A. Amato, arXiv:1101.1873v1 (2011).

\bibitem{Weibao} Wei Bao, Q. Huang, G. F. Chen, M. A. Green, D. M. Wang, J.
B. He, X. Q. Wang, Y. Qiu, arXiv:1102.0830 (2011).

\bibitem{Bin} Bin Zeng, Bing Shen, Genfu Chen, Jianbao He, Duming Wang,
Chunhong Li, Hai-Hu Wen, arXiv:1101.5117 (2011).

\bibitem{Tanatar} M. A. Tanatar, N. Ni, G. D. Samolyuk, S. L. Bud'ko, P. C.
Canfield, R. Prozorov, Phys. Rev. B 79, 134528 (2009).

\bibitem{Ruslan0} R. Prozorov and R.W. Giannetta, P. Fournier and R.L.
Greene, Phys. Rev. Lett. 85, 3700 (2000).

\bibitem{Martin} C. Martin, M. E. Tillman, H. Kim, M. A. Tanatar, S. K. Kim,
A. Kreyssig, R. T. Gordon, M. D. Vannette, S. Nandi, V. G. Kogan, S. L.
Bud'ko, P. C. Canfield, A. I. Goldman, and R. Prozorov

\bibitem{Ruslan1} R. Prozorov, M. A. Tanatar, Bing Shen, Peng Cheng, Hai-Hu
Wen, S. L. Bud'ko, and P. C. Canfield, Phys. Rev. B 82, 180513(R) (2010).

\bibitem{Ruslan2} R. Prozorov, M. E. Tillman, E. D. Mun, P. C. Canfield, New
Journal of Physics 11, 035004 (2009).

\bibitem{Ruslan3} R. Prozorov, N. Ni, M. A. Tanatar, V. G. Kogan, R. T.
Gordon, C. Martin, E. C. Blomberg, P. Prommapan, J. Q. Yan, S. L. Bud'ko,
and P. C. Canfield, Phys. Rev. B 78, 224506 (2008).

\bibitem{Tanatar2} M A Tanatar, N Ni, S L Bud'ko, P C Canfield and R
Prozorov, Supercond. Sci. Technol., 23 054002 (2010).

\bibitem{Cedomir} Hechang Lei, C. Petrovic, arXiv:1102.1010 (2011).
\end{thebibliography}
\end{document}